\documentclass[10pt,twocolumn,letterpaper]{article}

\usepackage[pagenumbers]{cvpr} % To force page numbers, e.g. for an arXiv version

\usepackage{booktabs}
\usepackage{times}
\usepackage{epsfig}
\usepackage{graphicx}
\usepackage{amsmath}
\usepackage{amssymb,amsfonts}
\usepackage{enumitem}
\usepackage{multirow}
\usepackage[table,xcdraw]{xcolor}
\usepackage{tabu}
\usepackage{url}
\usepackage{balance}

\usepackage[pagebackref,breaklinks,colorlinks]{hyperref}

\usepackage[capitalize]{cleveref}
\crefname{section}{Sec.}{Secs.}
\Crefname{section}{Section}{Sections}
\Crefname{table}{Table}{Tables}
\crefname{table}{Tab.}{Tabs.}

\def\cvprPaperID{14} % *** Enter the CVPR Paper ID here
\def\confName{CVPR}
\def\confYear{2022}

\begin{document}

\title{Learned Lossless Compression for JPEG via Frequency-Domain Prediction}

\author{
Jixiang~Luo, Shaohui~Li, Wenrui~Dai, Chenglin~Li, Junni~Zou, and Hongkai~Xiong\\
School of Electronic Information and Electrical Engineering, \\Shanghai Jiao Tong University, Shanghai, China\\
\{ljx123456, lishaohui, daiwenrui, lcl1985, zoujunni, xionghongkai\}@sjtu.edu.cn
}

\maketitle

%%%%%%%%% ABSTRACT
\begin{abstract}
JPEG images can be further compressed to enhance the storage and transmission of large-scale image datasets. Existing learned lossless compressors for RGB images cannot be well transferred to JPEG images due to the distinguishing distribution of DCT coefficients and raw pixels. In this paper, we propose a novel framework for learned lossless compression of JPEG images that achieves end-to-end optimized prediction of the distribution of decoded DCT coefficients. To enable learning in the frequency domain, DCT coefficients are partitioned into groups to utilize implicit local redundancy. An autoencoder-like architecture is designed based on the weight-shared blocks to realize entropy modeling of grouped DCT coefficients and independently compress the priors. 
We attempt to realize learned lossless compression of JPEG images in the frequency domain. Experimental results demonstrate that the proposed framework achieves superior or comparable performance in comparison to most recent lossless compressors with handcrafted context modeling for JPEG images. 
\end{abstract}

\begin{figure*}[!t]
\renewcommand{\baselinestretch}{1.0}
\centering
\includegraphics[width=\textwidth]{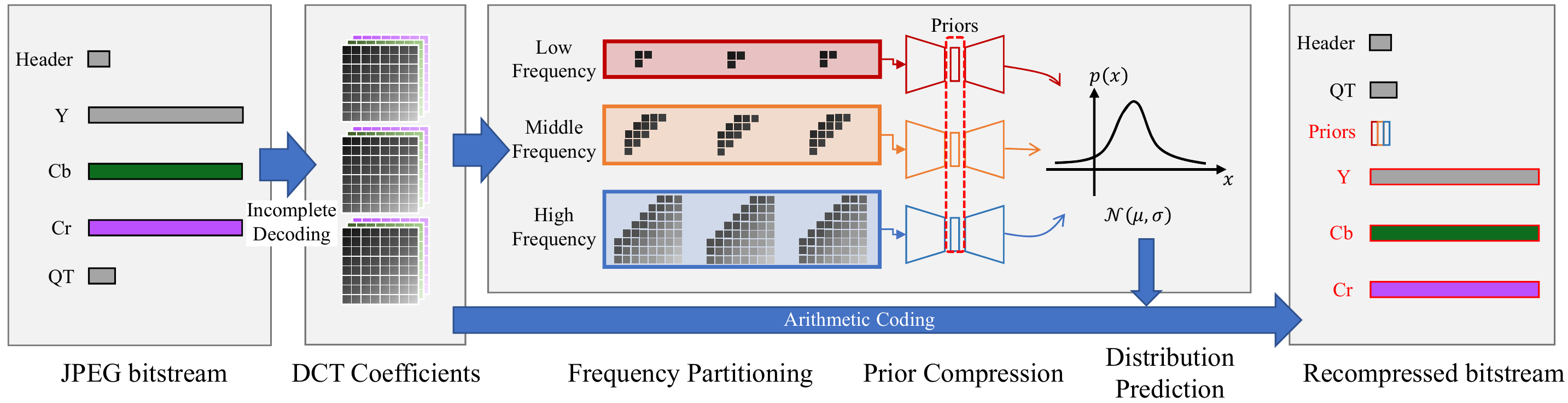}
\caption{Illustration of our proposed recompression framework. To implement a frequency domain recompression, JPEG bitstream is incompletely decoded to DCT coefficients. These DCT coefficients are further compressed with arithmetic coding, which benefits from a carefully designed distribution predictor. Besides, this distribution predictor compress the priors about the distributions as side information that contained in the recompressed bitstream.} \label{fig:data_flow}
\end{figure*}

%%%%%%%%% BODY TEXT
\section{Introduction}
Storage and transmission of large-scale image datasets, \emph{e.g.}, ImageNet~\cite{deng2009imagenet} and Flicker~\cite{flickr}, are necessary for training deep neural networks (DNNs). JPEG~\cite{wallace1991the} is the most popular image compression standard. The JPEG codec leverages transform coding, chroma subsampling, quantization and entropy coding in a sequence to remove spatial redundancies.
However, JPEG is inferior to JPEG2000\cite{rabbani2002jpeg2000} and BPG\cite{bpg} due to fixed $8\times 8$ discrete cosine transform (DCT) and Huffman coding. Recently, task-specific (lossy) and task-free (lossless) methods have been developed to further compress JPEG images.

Task-specific methods compress the JPEG images with the guidance of image processing tasks. Liu~\emph{et.~al}~\cite{liu2018deepn} developed DeepN-JPEG, a JPEG compression framework for image classification based on the high-frequency bias observed in experiments. DeepN-JPEG achieves about 350\% compression ratio on ImageNet without degrading the classification performance of deep neural networks. Li~\emph{et.~al}~\cite{li2020optimizing} optimized JPEG quantization table with sorted random search and composite heuristic optimization and achieves a gain of 20\%-200\% compression ratio at the same accuracy. Besides storage reduction, task-specific methods can also improve the performance of image processing. For example, Choi~\emph{et.~al}~\cite{choi2020task} estimated image-specific quantization tables with deep neural networks to improve the tasks of image classification performance, image captioning, and visual quality. However, task-specific compression methods are lossy, as they introduce extra distortion to the input JPEG images by adjusting the quantization tables.

Task-free methods are developed for universal compression of JPEG images. Lepton~\cite{horn2017design} is one of the representative work that utilizes manufactured context models for precise distribution prediction on each discrete cosine transformation (DCT) coefficient. Lepton provides about 23\% bitrate saving over original JPEG images and achieves efficient decompression via a parallelized arithmetic coding. %As a reliable method, it has been deployed on Dropbox file-storage backend for several years. 
%Other works that losslessly recompress existing JPEG files includes mozjpeg \cite{mozjpeg} and Brunsli \cite{brunsli}. They achieve around 10\% and 22\% file size reduction over JPEG images.
Besides Lepton, mozjpeg \cite{mozjpeg} and Brunsli \cite{brunsli} can also reduce the size of JPEG images by around 10\% and 22\% without introducing extra distortion, respectively. It is worth mentioning that universal lossless compression algorithms such as Lempel-Ziv-Markov chain-Algorithm (LZMA)~\cite{pavlov20197z} can also be employed on the JPEG images, but can only achieve a trivial compression gain, \emph{e.g.}, about 1\%. Although these lossless JPEG compressors are practical and efficient, they suffer from tedious design of context models. For example, Lepton uses kinds of manually designed contexts to model the conditional distribution and requires enormous statistical experiments to determine the parameters for prediction.

Recent development in end-to-end compression reveals the potential of getting rid of tedious handcrafted context modeling. The neural-network-based entropy models are developed for differentiable distribution modeling of the latent representations. However, existing methods for learned lossless compression are developed for RGB images and cannot be directly employed to effectively compress JPEG images (\emph{i.e.}, DCT coefficients). Recalling lossy image compression, the hyper-prior based entropy model~\cite{balle2018variational} can dramatically improve the rate-distortion performance of end-to-end image compression methods. In the hyper-prior model, the latent representations is supposed to follow certain parameterized distributions (\emph{e.g.} Gaussian distribution and Laplace distribution). Then the parameters for these distributions are predicted with a neural network, and utilized for arithmetic coding. To enhance the decoding process, the information about these distributions is compressed and transmitted to the decoder as a packed prior. 

Inspired by the hyper-prior model, in this paper, we propose a novel framework for lossless compression of JPEG images. As depicted in Figure~\ref{fig:data_flow}, the proposed framework achieves end-to-end optimized distribution prediction for arithmetic coding of DCT coefficients by incompletely decoding JPEG bitstream. The contributions of this paper are summarized as below. 
\begin{itemize}
\item We propose a novel framework for learned lossless compression of JPEG images. The proposed framework achieves comparable performance to the carefully designed traditional methods such as Lepton. 
\item We achieve end-to-end optimized distribution prediction of DCT coefficients incompletely decoded from the JPEG images via frequency partitioning and learning. Grouped DCT coefficients are adopted to improve the compression performance.
%We observe that compression with grouped coefficients can improve the compression performance.
\item We design the weight-shared residual blocks to constitute an autoencoder-like architecture that improves compression performance and maintains a low memory consumption during training.

\end{itemize}

This work is the learned lossless compressor specifically designed for JPEG images. Different from existing learned lossless methods, the DCT coefficients are partitioned into several frequency groups to enable end-to-end optimized distrubtion prediction. An autoencoder-like architecture is designed based on weight-shared blocks to realize entropy modeling of grouped DCT coefficients and independently compress the priors. Experimental results show that the proposed method outperforms LZMA, mozjpeg, and Brunsli, and is comparable to Lepton in terms of compression ratio. 

\section{Related Work}
\label{sec:related_work}
\subsection{Learned Lossless Compression}
Lossless compression has been studied for both universal data and images for a long time, and recent development of deep learning methods has stimulated new researches in this field. For example, DeepZip~\cite{goyal2018deepzip} used recurrent neural networks (RNNs) and bits-back coding~\cite{kingma2019bit} to realize lossless compression for universal data, especially for sequence data. Aiming at lossless image compression, L3C~\cite{mentzer2019practical} estimated the distribution of each pixel in RGB domain with a serialized hierarchical probabilistic model. Moreover, Mentzer~\emph{et.~al}~\cite{mentzer2020learning} and Cheng~\emph{et.~al}~\cite{cheng2020learned} suggested that compressing residual of compressed image with traditional methods is also feasible for lossless image compression with end-to-end models.

However, these models are not practical for JPEG image recompression for their low efficiency. The JPEG images have been lossy compressed and typically have a file size more than 20x smaller than the original image. But the most efficient lossless compression methods can only achieve 2x to 3x compression rate. Thus, the lossless compression in RGB domain cannot improve the compression rate of JPEG images. Therefore, a DCT domain compression is required.

 \begin{figure*}
    \centering
    \includegraphics[width=\textwidth, height=.40\textheight]{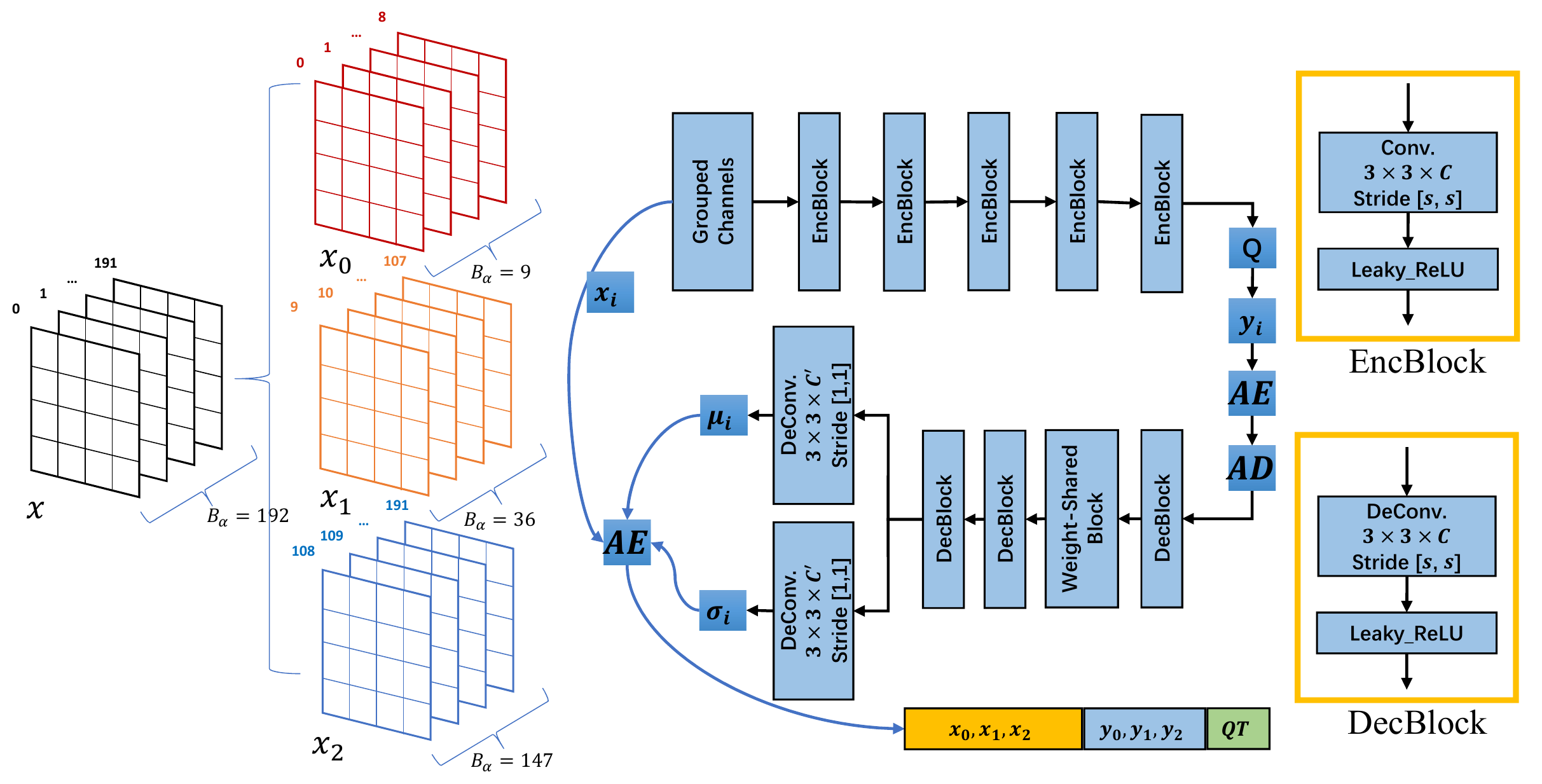}
    \caption{The left part is the data processing: $B_\alpha$ represents the number of channels. Then \textit{Frequency Band} is spilt into three \textit{Grouped Channels}. The right part is the network structure: The input is \textit{Grouped Channels}, namely $x_i$. \textit{Conv} and \textit{DeConv} is the convolution layer with $3\times 3$ kernel, $C$ channels and $[s,s]$ stride. Besides, the stride of first two \textit{EncBlock}s and the last two \textit{DecBlock}s are $[2, 2]$, the others are $[1, 1]$.  We set $C=48$ of the last \textit{EncBlock}, and the others are $384$.  \textit{Leaky\_ReLU} is the activation function. \textit{Q} is the quantization, $y_i$ is the features to be encoded. \textit{AE} and \textit{AD} are the arithmetic encoder and arithmetic decoder. $\mu_i$, $\sigma_i$ are the parameters for arithmetic coder. $C_i$ is the output channel and it is the same with input channels of grouped channels. Then $x_i$, $y_i$ and \textit{QT} are compressed to form the final bitstream. } 
    \label{fig:structure}
\end{figure*}

\subsection{Frequency Learning}
Conventional neural networks utilize RGB images as input, where the spatial information would be captured. For saving decoding time of JPEG images, methods exploiting DCT coefficients are explored. Gueguen~\emph{et.~al}~\cite{NEURIPS2018_7af6266c} trained a convolutional neural network (CNN) directly on the DCT coefficient acquired from JPEG bitstream, which gains acceleration compared with standard residual network (ResNet). Ehrlich~\emph{et.~al}~\cite{Ehrlich_2019_ICCV} redefined convolution, batch normalization, and ReLU leveraging the linearity of the JPEG transform. Xu~\emph{et.~al}~\cite{learningfrequency} used DCT coefficients as input and selected the most significant components to perform image inference, which reduce the burden of data transmission. This method is verified in image detection and classification, with a superior performance over conventional methods. These methods proves the potential of frequency learning and inspire us to use end-to-end models for lossless frequency compression.

\section{Methodology}
\label{sec:method}
This section demonstrates the framework for lossless compression of JPEG images. We first present the pipeline of the proposed framework, and then describe implementation details of the end-to-end distribution predictor.

\subsection{Proposed Framework}
The proposed framework is developed for the DCT coefficients obtained by incomplete decoding of JPEG images. For clarity, we denote the tensor of DCT coefficients as $x \in \mathbb{R}^{H\times W\times 192}$, that consists of 64 frequency components of DCT coefficients over 3 color planes. The details of the arrangement is described in Section~\ref{sec:detail}. As shown in Figure~\ref{fig:structure}, $x$ is partitioned into several frequency groups in the sense of ``low frequency", ``middle frequency", and ``high frequency". The partitioned groups are denoted as $x_0, x_1, \cdots, x_n$ with $x_i \in \mathbb{R}^{H\times  W\times C_i}$ and $\sum_i C_i = 192$. 
$x_i$ is supposed to obey a multivariate Gaussian distribution, where the mean $\mu_i$ and the scale $\sigma_i$ are predicted with distribution predictor.
Moreover, the distributions of each group are estimated separately with a specific distribution predictor. The predictors share the same structure, while the parameters are learned independently since the scale and correlation of different frequency is distinguished. Overall, the predictors are constructed like an auto-encoder, where the output is the means and scales for Gaussian distribution. Here, the encoder and decoder are denoted as $\mathcal{E}_i$ and $\mathcal{D}_i$.
\begin{equation}
\begin{aligned}
y_i= \lfloor\mathcal{E}_i(x_i)\rceil, \quad
[\mu_i, \sigma_i]= \mathcal{D}_i(y_i),
\end{aligned}
\end{equation}
where $\lfloor \cdot \rceil$ represents the rounding operation and $y_i$ is the quantized output of the encoder. $y_i$ is the prior of $\mu_i$ and $\sigma_i$, and is quantized to integer symbols for entropy coding. $y_i$ is encoded and embedded into the bitstream, and is transmitted to the decoder.

Considering end-to-end training for the distribution predictors, the whole framework is optimized based on a joint loss to balance the performance over all frequency components. The loss function is designed as
\begin{align}
\label{eq:loss}
L &= \sum_i (R_{x_i}  + \lambda R_{y_i}) \nonumber\\
&= \sum_i\left(\mathbb{E}[-\log_2(p_{x_i})] + \lambda \mathbb{E}[-\log_2(p_{y_i})]\right),
\end{align}
where $R_{x_i}$ and $R_{y_i}$ is the average bit consumption of $x_i$ and $y_i$. The probability of $x_i$ can be inferred from the parameterized Gaussian distritbution. With the estimated $\mu_i$ and $\sigma_i$, the probability of integer symbol $a = x_i^{h,w}$ in $x_i$ is
\begin{equation}
p(a, \sigma, \mu) = \int_{a-0.5}^{a+0.5} \frac{1}{\sqrt{2\pi}\sigma} e^{-\frac{(a - \mu)^2}{2\sigma^2}},
\end{equation}
where $\mu = \mu_i^{h,w}$ and $\sigma = \sigma_i^{h,w}$ are the mean and scale on the corresponding positions of $\mu_i$ and $\sigma_i$. The probability of $y_i$ is estimated with the method introduced in \cite{balle2016end}. It is worthy mentioning that $\lambda$ in Equation~\eqref{eq:loss} is expected to be 1 for the lossless compression task, but we find that progressively increasing $\lambda$ would improve the compression performance. The details of the tuning strategy is elaborated in Section~\ref{sec:trainig_strategies}.

\subsection{Implementation Details}\label{sec:detail}
\begin{figure}
    \centering
    \includegraphics[width=.47\textwidth]{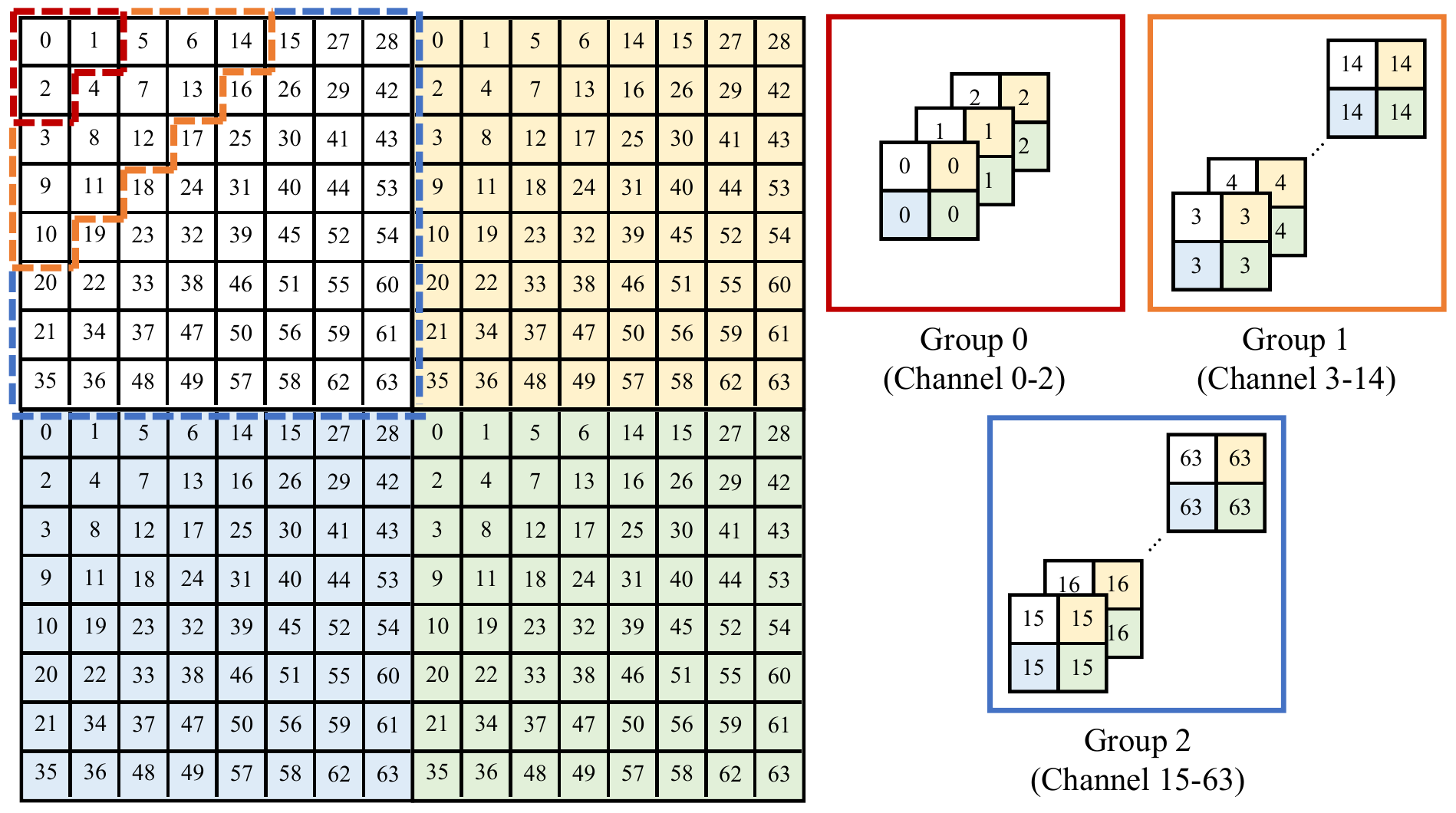}
    \caption{An example of frequency partitioning proposed in this paper. The left of this figure shows four adjacent blocks of $8\times 8$ DCT coefficients of luminance (Y plane), where the indices of coefficients follow the Zig-Zag scanning of JPEG. In this example, the 64 coefficients are partitioned into 3 groups according to frequency. In each group, the spatial position of the adjacent blocks is maintained, while coefficients of different coefficients is arranged across channels. Thus, the three groups in this figure have the shapes   $2\times 2\times 3$, $2\times 2\times 12$, and $2\times 2\times 49$.}
    \label{fig:freq_part}
\end{figure}

\textbf{Arrangement of DCT Coefficients .} 
Suppose the original image has a size of $N\times M$, then these coefficients are stored in $\lceil\frac{N}{8}\rceil\times \lceil\frac{M}{8}\rceil$ blocks of $8\times 8$, where $\lceil \cdot \rceil$ means the ceil operation. Besides, considering the three color planes (\emph{i.e.}, Y, Cb, and Cr) adopted in JPEG for compressing color image, $x$ has the shape of $\lceil\frac{N}{8}\rceil\times \lceil\frac{M}{8}\rceil\times 8\times 8\times 3$. To simplify the notation of this 5-dimensional tensor, we rearrange the coefficients according to their frequency and color planes, while keep the first two dimensions unchanged. As shown in Figure~\ref{fig:freq_part}, the $8\times 8$ DCT coefficients in each block is indexed in a Zig-Zag order ranging from 0 to 63. The coefficients on different color planes with the same index placed in the following order: Y$\rightarrow$Cb$\rightarrow$Cr$\rightarrow$Y$\rightarrow$Cb$\rightarrow$Cr$\cdots$. Moreover, we define $H \equiv \lceil\frac{N}{8}\rceil, W \equiv \lceil\frac{M}{8}\rceil$. Then the 5-dimensional tensor $x$ is reshaped into the shape mentioned before, as $H\times W\times 192$.
 
\textbf{Network Architecture}. The main body is the autoencoder as shown in Figure~\ref{fig:structure}, including encoder and decoder. Encoder consists of  five \textit{EncBlock}, and each \textit{EncBlock} is composed of convolutional layer with $3 \times 3$  kernel and leaky ReLU activation function. The first two \textit{EncBlock} have stride with $2$, thus encoder subsamples with $4$ times to transfer input coefficients $x_i$ into $\dot{y}_i$. Then we quantize $\dot{y}_i$ into $y_i$ for arithmetic encoding and decoding. And decoder has \textit{DecBlock} and \textit{Weight-Shared Block}, and each \textit{DecBlock} is composed of deconvolutional layer with $3 \times 3$  kernel and leaky ReLU activation function. The last two \textit{DecBlock} has stride with $2$ and decoder upsamples the extracted feature with $4$ time. Then the last convolutional layers output the parameters of Gaussian $\mu_i$ and $\sigma_i$ to generate the probability of input $x_i$. 

\begin{figure}
    \centering
    \includegraphics[width=.48\textwidth]{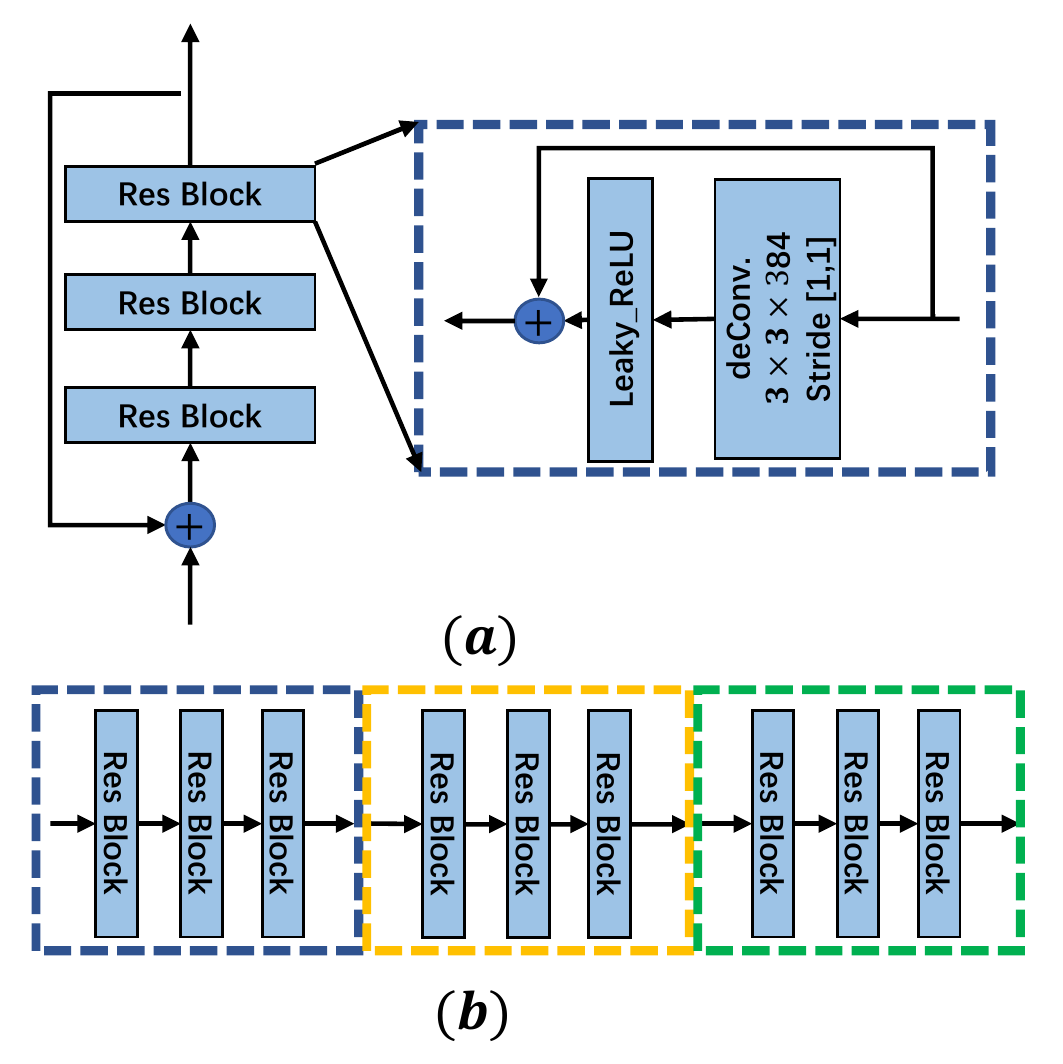}
    \caption{The structure of \textit{Weight-Shared Block}. Then \textit{Res\_Block} includes \textit{deConv} and \textit{Leaky\_ReLU}. And \textit{deConv} is the convolution layer with $3\times3$ kernel, $384$ channels and $[1, 1]$ stride. \textit{Leaky\_ReLU} is the activation function. $+$ mean addition element by element. Then $a$ is the wieght-shared network, $b$ is the serial connection of three $Res_Block$. When $a$ has three iterations, $a$ has the same structure with $b$, but three blocks in $a$ share the same parameters.}
    \label{fig:res_block}
\end{figure}

\textbf{Weight-Shared Block.} The weight-shared block in Figure~\ref{fig:res_block} is introduced in the decoder side to facilitate training. When Figure~\ref{fig:res_block}($a$) has three iterations, Figure~\ref{fig:res_block}($a$) has the same structure with Figure~\ref{fig:res_block}($b$). But three blocks in Figure~\ref{fig:res_block}($a$) share the same parameters. Different from the ResNet in \cite{he2016deep}, the weight of these blocks share the same weights. Thus we name it weight-shared block. Besides, it has forward flow(from input to output) as shown in the left of  Figure~\ref{fig:res_block}($a$) and the backward flow(from output to input) as shown in the right of  Figure~\ref{fig:res_block}($a$) at the same time.  It reduces the complexity of decoder, and improve the compression gain. 

\section{Frequency Partitioning}
\label{sec:freq_part}
In this section, we further validate the efficiency of frequency partitioning in the proposed framework. 

\subsection{Correlation Across Frequencies}\label{channel-corr}
\begin{table}[]
\resizebox{.48\textwidth}{!}{
\centering
\begin{tabular}{c|c|c|c|c|c}
\hline \hline
channels                    & bpsp                     & channels                    & bpsp                                            & channels   & bpsp            \\ \hline 
                            &                          &                             & {\color[HTML]{C8161E} }                         & {[}0, 3)   & 0.1227          \\ \cline{5-6} 
                            &                          & \multirow{-2}{*}{{[}0, 9)}  & \multirow{-2}{*}{\textbf{0.2559}} & {[}3, 9)   & 0.1481          \\ \cline{3-6} 
                            &                          &                             & {\color[HTML]{C8161E} }                         & {[}9, 18)  & 0.1822          \\ \cline{5-6} 
                            &                          &                             & {\color[HTML]{C8161E} }                         & {[}18, 30) & 0.1937          \\ \cline{5-6} 
\multirow{-5}{*}{{[}0, 45)} & \multirow{-5}{*}{0.8186} & \multirow{-3}{*}{{[}9, 45)} & \multirow{-3}{*}{\textbf{0.5453}} & {[}30, 45) & 0.1810           \\ \hline \hline
                           & 0.8186          &                             & \textbf{0.8012}                                 &            &  0.8277 \\ \hline
\end{tabular}}
\caption{The bits for grouped channels of kodim01. The number of  \textit{channels} $[m, n)$ is the index of $192$ frequency bands from $m$ to $n$. And \textit{bpsp} is the bit per sub-pixel. The last line is the sum of each component.}\label{channel-bpsp}
\end{table}

First, we compare the compression performance with three grouping strategies. The first 45 channels in the arranged DCT coefficients $x$ are utilized in this experiment. As shown in Table~\ref{channel-bpsp}, the 45 channels are split into 1/2/5 groups at different refinement levels. The best compression performance is achieved when the 45 channels are split into 2 groups. The reasons for that are of two aspects. First, the DCT coefficients are locally correlated. Second, the average intensities of different channels are distinguished. For example, the average intensity of channel 0 (DC) can be hundreds of times higher than channel 30-45, which may hinder the network from finding an optimum.

\begin{figure}
    \centering
    \includegraphics[width=.47\textwidth]{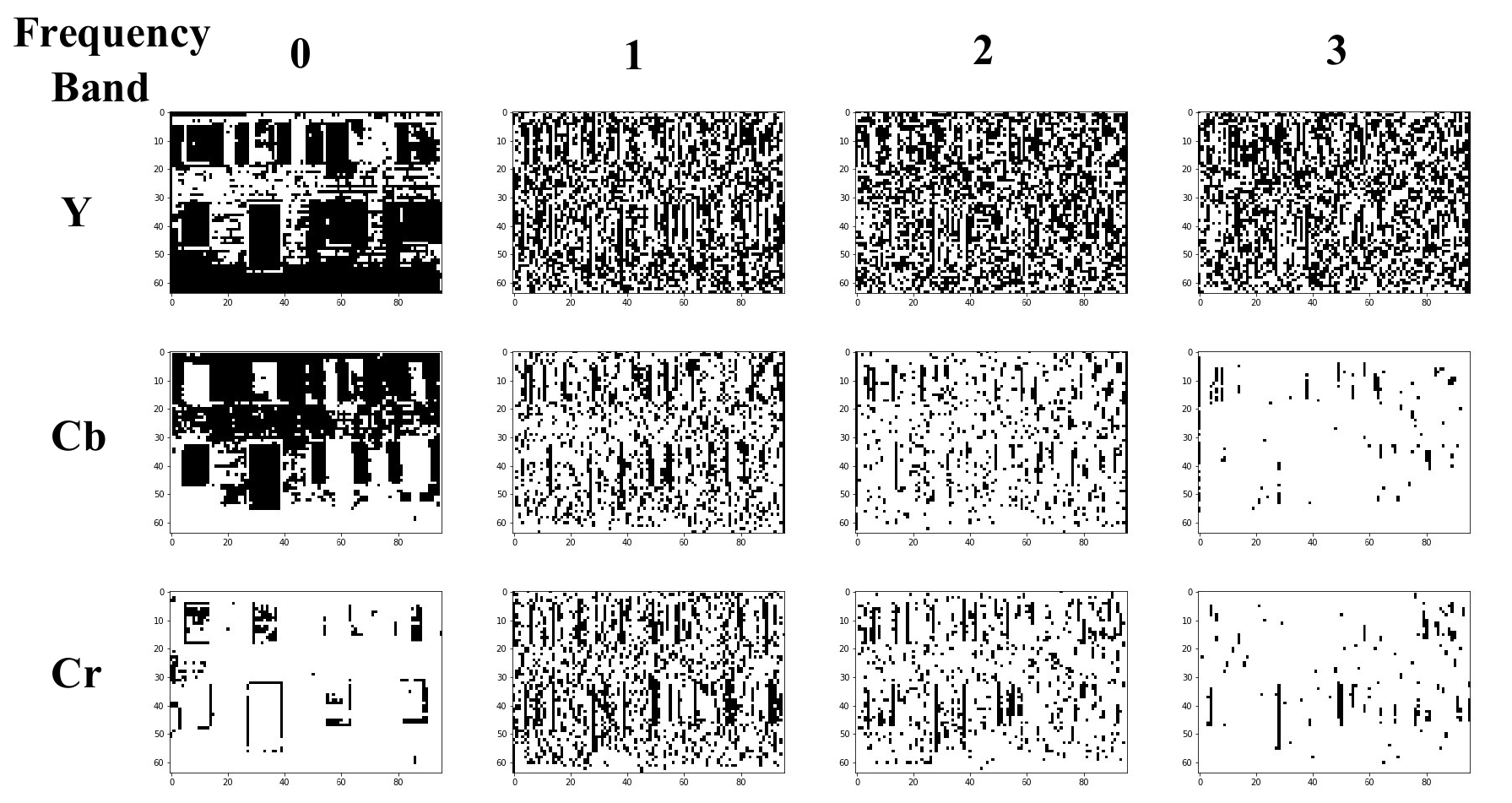}
    \caption{The most significant bit map of channel 0-3 of \textit{Y}, \textit{Cb} and \textit{Cr} planes.}
    \label{fig:frequency_band}
\end{figure}

Moreover, we visualize the DCT coefficients with binary map of Y, Cb, Cr color planes to depict the correlation among low frequency components in Figure~\ref{fig:frequency_band}. Figure~\ref{fig:frequency_band} is the most significant bit (MSB) map of 0-3 channels of the first 45 channels, which suggests high correlation in channel 0 across different color planes and some implicit correlation among other low frequency channels.  

\begin{figure}
    \centering
    \includegraphics[width=.49\textwidth]{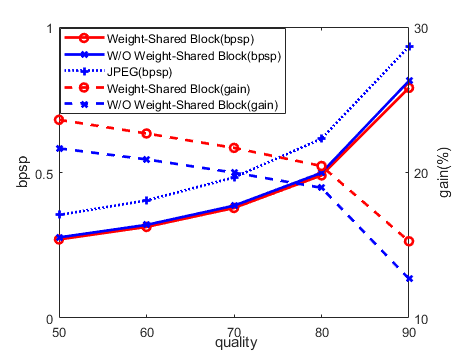}
    \caption{The ablation experiment of Weight-Shared Block on Kodak. The left y axis is \textit{bpsp}, and the right is the compression ratio. Then red lines are the model with weight-shared block, while the blue dotted lines are the model without weight-shared block. }
    \label{fig:weight-shared}
\end{figure}

\subsection{Frequency Partitioning Strategy}

Based on the experiments and channel correlation introduced in Section~\ref{channel-corr}, we split the $192$ channels into three groups. To illustrate the physical meaning of such partitioning strategy, we illustrate it in the original block DCT domain in Figure~\ref{fig:freq_part}. For gray-scale image with only luminance (Y) plane, the 64 channels that indexed in Zig-Zag order are split into 3 groups: $[0, 3), [3, 15), [15, 64)$. As for color images, the channels are grouped to $[0, 9), [9, 45), [45, 192)$, where the color planes are ordered as Y$\rightarrow$Cb$\rightarrow$Cr$\rightarrow$Y$\rightarrow$Cb$\rightarrow$Cr$\cdots$.

\begin{table*}[!t]
\centering
\resizebox{\textwidth}{!}{
\begin{tabular}{c|c|c|c|c|c|c|c|c}
\hline\hline
 \multicolumn{2}{c|}{Methods}  & Kodak\_50     & Kodak\_60     & Kodak\_70     & Kodak\_80     & Kodak\_90     &                                                                                       &                                                                                       \\ \cline{1-7}
Baseline                                                                           & JPEG    & 0.354         & 0.404         & 0.483         & 0.616         & 0.933         & \multirow{-2}{*}{\begin{tabular}[c]{@{}c@{}}Average Gain \\ to JPEG(\%)\end{tabular}} & \multirow{-2}{*}{\begin{tabular}[c]{@{}c@{}}Average Gain \\ to Others(\%)\end{tabular}} \\ \hline
Learned & proposed & \textbf{0.270(23.622)} & \textbf{0.313(22.682)} & \textbf{0.378(21.691)} & \textbf{0.490(20.447)} & \textbf{0.791(15.275)} & \textbf{20.743}                                                                                & -                                                                                     \\ \hline
  & LZMA    & 0.345(2.344)  & 0.398(1.462)  & 0.480(0.553)  & 0.617(-0.162) & 0.939(-0.661) & 0.707                                                                                 & {\color[HTML]{C00000} +20.036}                                                        \\ \cline{2-9} 
                                                                                   & mozjpeg & 0.324(8.287)  & 0.377(6.825)  & 0.456(5.600)  & 0.586(4.844)  & 0.881(5.553)  & 6.222                                                                                 & {\color[HTML]{C00000} +14.521}                                                        \\ \cline{2-9} 
                                                                                   & Brunsli & 0.271(23.272) & 0.317(21.681) & 0.385(20.220) & 0.500(18.903) & 0.761(18.479) & 20.511                                                                                & {\color[HTML]{C00000} +0.232}                                                         \\ \cline{2-9} 
\multirow{-4}{*}{\begin{tabular}[c]{@{}c@{}}Handcrafted  \\  Methods\end{tabular}} & Lepton  & 0.265(25.159) & 0.309(23.564) & 0.377(21.972) & 0.491(20.286) & 0.756(18.931) & 21.982                                                                                & {\color[HTML]{00B050} -1.249}                                                         \\ \hline
\end{tabular}}
\caption{Compression performance on Kodak by our method, baseline (JPEG) methods, and existing handcrafted methods in terms of bits per sub-pixel (bpsp). We emphasize the difference in percentage to our approach for each other method in red if our method outperforms the other method and in green otherwise. For \textit{Kodak\_m}, $m$ is the quality factor to control the quality of JPEG images. }\label{result_kodak}
\end{table*}

\begin{table*}[!t]
\centering
\resizebox{\textwidth}{!}{
\begin{tabular}{c|c|c|c|c|c|c|c|c}
\hline\hline
 \multicolumn{2}{c|}{Methods} & Set5\_50              & Set5\_60              & Set5\_70              & Set5\_80              & Set5\_90              &                                                                                       &                                                                                       \\ \cline{1-7}
Baseline                                                                           & JPEG    & 0.454                  & 0.516                  & 0.611                  & 0.767                  & 1.108                  & \multirow{-2}{*}{\begin{tabular}[c]{@{}c@{}}Average Gain \\ to JPEG (\%)\end{tabular}} & \multirow{-2}{*}{\begin{tabular}[c]{@{}c@{}}Average Gain \\ to Others (\%)\end{tabular}} \\ \hline
Learned & proposed & \textbf{0.343(24.449)} & \textbf{0.391(24.258)} & \textbf{0.467(23.472)} & \textbf{0.596(22.271)} & \textbf{0.918(17.153)} & 22.321                                                                                & -                                                                                     \\ \hline
                                                                                   & lzma    & 0.450(0.871)           & 0.514(0.476)           & 0.610(0.140)           & 0.767(-0.101)          & 1.112(-0.330)          & 0.211                                                                                 & {\color[HTML]{C00000} +22.11}                                                         \\ \cline{2-9} 
                                                                                   & mozjpeg & 0.434(4.507)           & 0.494(4.348)           & 0.582(4.632)           & 0.724(5.612)           & 1.023(7.636)           & 5.347                                                                                 & {\color[HTML]{C00000} +16.974}                                                        \\ \cline{2-9} 
                                                                                   & Brunsli & 0.362(20.202)          & 0.413(19.95)           & 0.516(19.417)          & 0.615(19.76)           & 0.882(20.388)          & 19.943                                                                                & {\color[HTML]{C00000} +2.378}                                                         \\ \cline{2-9} 
\multirow{-4}{*}{\begin{tabular}[c]{@{}c@{}}Handcrafted  \\  Methods\end{tabular}} & Lepton  & 0.354(20.055)          & 0.405(21.628)          & 0.481(21.184)          & 0.606(20.946)          & 0.874(21.108)          & 20.984                                                                                & {\color[HTML]{C00000} +1.437}                                                         \\ \hline

\end{tabular}}
\caption{Compression performance on Set5 by our method, baseline(JPEG), and existing handcrafted methods in terms of bits per sub-pixel (bpsp). We emphasize the difference in percentage to our approach for each other method in red if our method outperforms the other method and in green otherwise. For \textit{Set\_m}, $m$ is the quality factor to control the quality of JPEG images. }\label{result-set5}
\end{table*}

\begin{figure*}
    \centering
    \includegraphics[width=\textwidth]{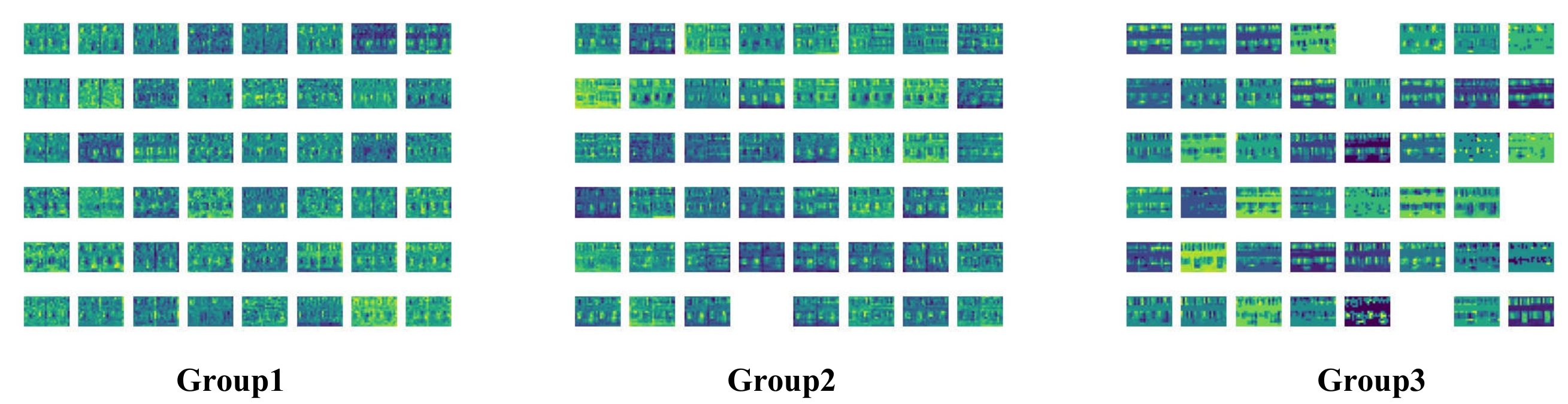}
    \caption{The visualization of features from frequency learning for the JPEG format of kodim01. The blank in group $2$ and $3$ is the redundant channel, which can not learn from input.}
    \label{fig:features}
\end{figure*}

\section{Experiments}
\label{sec:experiment}

\subsection{Dataset}
To train and test our neural network, we process the training and testing data according to the Figure~\ref{fig:data_flow}. Moreover, We must obtain DCT coefficients from JPEG images, here we follow the code\footnote{\url{https://github.com/dwgoon/jpegio}} to get the quantization table and coefficients. Then the training dataset is DCT coefficients of Flickr data\cite{flickr}. Firstly we crop these JPEG images with $
256 \times 256$, and then the input is $32 \times 32$ since the DCT is $8 \times 8$. Besides the training image batch is $4$. And the testing dataset is Kodak\cite{kodak} and Set5\cite{bevilacqua2012low}. We just transfer lossless format PNG as JPEG by controlling the quality factor with library "jpeg-9c"\footnote{\url{https://www.ijg.org/}}. The set of quality factor is ${50, 60, 70, 80, 90}$, and $50$ means the worst quality and $90$ otherwise. It covers the commonly used quality for JPEG. Moreover the main memory cost of cloud and local machine is mainly from the JPEG images with high quality. And then we process testing data with the method same as the training data. 

\subsection{Training Strategies}
\label{sec:trainig_strategies}
We utilize Adam\cite{kingma2014adam} optimizer to train our whole network with one GPU for two days. From the loss function in Equation~\eqref{eq:loss}, we train our neural network by balancing the rate of DCT coefficients and extracted features. To obtain higher compression ratio, we adjust the vaule of $\lambda$ during training instead of a constant value. Firstly we set small value for it, like $0.001$, then when training at 1M steps, we set a bit higher value, like $0.01$. At last, we set $\lambda=1$ for the last 1M step.  Meanwhile, we adjust the learning rate by the exponential decay with $0.9$ until 10M steps.

\subsection{Visualization of Features}
To verify the frequency learning for lossless compression, we visualize the features that neural network learn from DCT coefficients. From the Figure~\ref{fig:features}, different frequency bands obtain different features. Neural network learns about the smoothed pixels from low frequecny as in \textit{Group1} in Figure~\ref{fig:features} and learns about dramatically changing pixels as in \textit{Group3}, like the edge and contours. Besides, since larger part of high frequency of DCT coefficients is zero, there exist some redundant feature channels in \textit{Group2} and \textit{Group3}.  It is the truth that the neural network can learn form DCT coefficients as the original RGB images, since DCT is reversible. Namely, CNN can learn from the inverse DCT. So we can infer the learning process, firstly it transforms the DCT coefficients into RGB images, and then learn the image pattern as the previous network. More importantly, neural network can learn from part frequency bands instead of all DCT coefficients. Since here is the lossless compression, we must process all frequency bands.

\begin{figure}
    \centering
    \includegraphics[width=.49\textwidth]{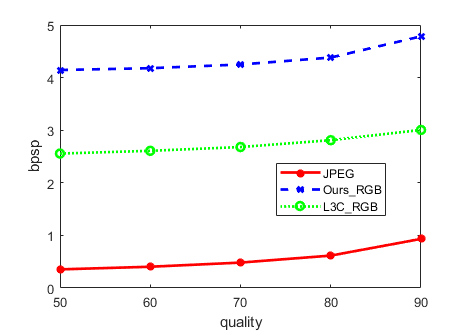}
    \caption{Learned methods with RGB input for JPEG images. We use the pretrained model L3C with green line. We use the learning methods with the same structure of Figure\ref{fig:structure} at blue line. The last red line is the baseline of JPEG image. }
    \label{fig:dctVSrgb}
\end{figure}

\subsection{Results}
We achieve more than 20\% compression gain for JPEG images, as shown in Tables~\ref{result_kodak} and~\ref{result-set5}. Our compression performance outperforms \textit{LZMA} about $20\%$ and \textit{mozjpeg} about $15\%$, and it is comparable to \textit{Brunsli} and \textit{Lepton}.  Besides, our method even has higher compression  gain for the dataset \textit{Kodak\_80} with $20.447\%$ to $20.286\%$. And we have higher compression ratio on Set5 dataset even for Brunsli and Lepton with $2.378\%$ and $1.437\%$ gain. 

\subsection{Ablation Studies}
\subsubsection{DCT vs. RGB}
To verify the efficiency of learning in frequrency domain, we conduct the experiments with the RGB input. And Figure~\ref{fig:dctVSrgb} shows all leaning in RGB has higher bpsp than the original images. Namely, they have no compression gain for JPEG images, though L3C\footnote{\url{https://github.com/fab-jul/L3C-PyTorch/}} has better performance in image with lossless format (BPG, JPEG2000 and so on). Besides we train our structure with the JPEG images, and test on Kodak dataset, its performance is far more worse than JPEG. To achieve lossless compression, neural network learn from each pixel and allocate its probability. Thus neural network can not identify whether the input image is lossless format. And that is our key motivation for learning in frequency domain.  

\subsubsection{Number of Groups}
\begin{figure}
    \centering
    \includegraphics[width=.46\textwidth]{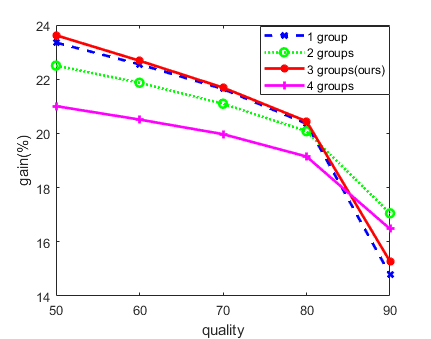}
    \caption{Compression gain of models with $1, 2, 3, 4$ groups, respectively.}
    \label{fig:groups}
\end{figure}

To explore the variable about the number of group, we set the following experiments. We split the whole $192$ channels into $1, 2, 3, 4$ groups. When it is $1$ group, we have the whole frequency bands $[0, 192)$ to jointly process these DCT coefficients. And experiments with $2$ groups has $[0, 9), [9, 192)$ frequency bands, and $3$ groups has  $[0, 9), [9, 45), [45, 192)$ , and $4$ groups has  $[0, 9), [9, 45), [45, 108), [108, 192)$, respectively. We split these groups according to the Zig-Zag order of $8\times8$ DCT, and it corresponds to super low,  low, middle and high frequency. From the Figure~\ref{fig:groups} a single model can not achieve best performance for all qualities. More groups also do not have better compression ratio as the model with $4$ groups is worse than model with $2$ group for all qualities. To be mentioned, model with $2$ groups has the best performance at quality $90$ with regard to other models. We finally use model with three groups for testing because it has the best average performance.

\begin{table}[]
    \centering
    \begin{tabular}{c|c|c|c|c}
    \hline\hline
     Group Number & 1 & 2 & 3 & 4  \\ \hline \hline
     Kodak\_50 & 0.271 &0.274 & 0.270 & 0.279 \\ \hline
     Kodak\_60 & 0.313 &0.316 & 0.313 & 0.321 \\ \hline
     Kodak\_70 & 0.378 &0.381 & 0.378 & 0.386 \\ \hline
     Kodak\_80 & 0.490 &0.492 & 0.490 & 0.498 \\ \hline
     Kodak\_90 & 0.795 &0.774 & 0.791 & 0.779 \\ \hline
    %  Model Size(MB) & 211.7 &352.2 & 488.8 & 628.1 \\ \hline
    \end{tabular}
    \caption{Compression performance about bpsp of models with $1, 2, 3, 4$ groups, respectively. }
    \label{tab:model_size}
\end{table}

\subsubsection{Weight-Shared Block}
We conduct the experiments to verify the effectiveness of weight-shared block and the results are shown in Figure~\ref{fig:weight-shared}. Specifically, we disconnect the back flow in Figure~\ref{fig:res_block}($a$) to train the whole network with the same configuration. And it actually improves the performance by 2\% for images across all qualities. 

\section{Conclusion and Discussion}
\label{sec:conclusion}
As far as we know, we are the first to utilize deep learning method on frequency domain for lossless compression. 
The DCT coefficients are partitioned into several groups for efficient compression, which is based on the observation that DCT coefficients have implicit local correlation. Thus, joint processing adjacent channels can improve the lossless compression performance. Besides, different from the autoencoder with the symmetrical structure, we introduce the extra module for the decoder, such as weight-shared block, because the pattern of DCT coefficient is hard to capture, especially the altering coefficients. Finally we achieve the comparable performance to other traditional non-learned methods. Different from the rate-distortion loss $R+\lambda D$ for lossy compression, lossless compression has no distortion. In this paper, we just optimize the rate of input DCT coefficients and the extracted priors. However we also introduce the Lagrange factor $\lambda$ to balance the rate of each part. 

{\small
\bibliographystyle{ieee_fullname}
\bibliography{ms}

\begin{thebibliography}{10}\itemsep=-1pt

\bibitem{bsds}
The berkeley segmentation dataset and benchmark.
\newblock
  \url{https://www2.eecs.berkeley.edu/Research/Projects/CS/vision/bsds/}.

\bibitem{brunsli}
Brunsli - github.
\newblock {\em \url{https://github.com/google/brunsli}}.

\bibitem{kodak}
Kodak lossless true color image suite (pho-tocd pcd0992).
\newblock \url{http://www.tp-ontrol.hu/index.php/TP_Toolbox}.

\bibitem{flickr}
Mirflickr.
\newblock \url{http://press.liacs.nl/mirflickr/mirflickr1m/}.

\bibitem{mozjpeg}
Mozilla mozjpeg.
\newblock {\em
  \url{https://blog.mozilla.org/research/2014/03/05/introducing-the-mozjpeg-project/}}.

\bibitem{Alpher02}
Alvin Alpher.
\newblock Frobnication.
\newblock {\em Journal of Foo}, 12(1):234--778, 2002.

\bibitem{Alpher03}
Alvin Alpher and Ferris P.~N. Fotheringham-Smythe.
\newblock Frobnication revisited.
\newblock {\em Journal of Foo}, 13(1):234--778, 2003.

\bibitem{Alpher04}
Alvin Alpher, Ferris P.~N. Fotheringham-Smythe, and Gavin Gamow.
\newblock Can a machine frobnicate?
\newblock {\em Journal of Foo}, 14(1):234--778, 2004.

\bibitem{ayyoubzadeh2020lossless}
Seyed~Mehdi Ayyoubzadeh and Xiaolin Wu.
\newblock Lossless compression of mosaic images with convolutional neural
  network prediction.
\newblock {\em arXiv preprint arXiv:2001.10484}, 2020.

\bibitem{balle2016end}
J. Ball{\'e}, V. Laparra, and E.~P. Simoncelli.
\newblock End-to-end optimized image compression.
\newblock In {\em Proceedings of the 5th International Conference on Learning
  Representations}, Toulon, France, April 2017.

\bibitem{balle2018variational}
J. Ball{\'e}, D. Minnen, S. Singh, S.~J. Hwang, and N. Johnston.
\newblock Variational image compression with a scale hyperprior.
\newblock In {\em Proceedings of the 6th International Conference on Learning
  Representations}, Vancouver, BC, Canada, April 2018.

\bibitem{bpg}
F. Bellard.
\newblock {BPG} image format.
\newblock \url{https://bellard.org/bpg/}.

\bibitem{bevilacqua2012low}
Marco Bevilacqua, Aline Roumy, Christine Guillemot, and Marie~Line
  Alberi-Morel.
\newblock Low-complexity single-image super-resolution based on nonnegative
  neighbor embedding.
\newblock 2012.

\bibitem{blaszczyk2012paq}
Krzysztof Blaszczyk, Peter Rossmanith, Dipl-Inf~Alexander Langer, and
  Dipl-Inf~Felix Reidl.
\newblock Paq compression algorithm.
\newblock 2012.

\bibitem{cao2020lossless}
Sheng Cao, Chao-Yuan Wu, and Philipp Kr{\"a}henb{\"u}hl.
\newblock Lossless image compression through super-resolution.
\newblock {\em arXiv preprint arXiv:2004.02872}, 2020.

\bibitem{chen2019drop}
Yunpeng Chen, Haoqi Fan, Bing Xu, Zhicheng Yan, Yannis Kalantidis, Marcus
  Rohrbach, Shuicheng Yan, and Jiashi Feng.
\newblock Drop an octave: Reducing spatial redundancy in convolutional neural
  networks with octave convolution.
\newblock In {\em Proceedings of the IEEE International Conference on Computer
  Vision}, pages 3435--3444, 2019.

\bibitem{cheng2020learned}
Zhengxue Cheng, Heming Sun, Masaru Takeuchi, and Jiro Katto.
\newblock Learned lossless image compression with a hyperprior and discretized
  gaussian mixture likelihoods.
\newblock In {\em ICASSP 2020-2020 IEEE International Conference on Acoustics,
  Speech and Signal Processing (ICASSP)}, pages 2158--2162. IEEE, 2020.

\bibitem{choi2020task}
Jinyoung Choi and Bohyung Han.
\newblock Task-aware quantization network for jpeg image compression.
\newblock In {\em European Conference on Computer Vision}, pages 309--324.
  Springer, 2020.

\bibitem{deng2009imagenet}
Jia Deng, Wei Dong, Richard Socher, Li-Jia Li, Kai Li, and Li Fei-Fei.
\newblock Imagenet: A large-scale hierarchical image database.
\newblock In {\em 2009 IEEE conference on computer vision and pattern
  recognition}, pages 248--255. Ieee, 2009.

\bibitem{ehrlich2020quantization}
Max Ehrlich, Larry Davis, Ser-Nam Lim, and Abhinav Shrivastava.
\newblock Quantization guided jpeg artifact correction.
\newblock In {\em Proceedings of the European Conference on Computer Vision}.
  Springer, 2020.

\bibitem{Ehrlich_2019_ICCV}
Max Ehrlich and Larry~S. Davis.
\newblock Deep residual learning in the jpeg transform domain.
\newblock In {\em Proceedings of the IEEE/CVF International Conference on
  Computer Vision (ICCV)}, October 2019.

\bibitem{gers1999learning}
Felix~A Gers, J{\"u}rgen Schmidhuber, and Fred Cummins.
\newblock Learning to forget: Continual prediction with lstm.
\newblock 1999.

\bibitem{goyal2018deepzip}
Mohit Goyal, Kedar Tatwawadi, Shubham Chandak, and Idoia Ochoa.
\newblock Deepzip: Lossless data compression using recurrent neural networks.
\newblock {\em arXiv preprint arXiv:1811.08162}, 2018.

\bibitem{NEURIPS2018_7af6266c}
Lionel Gueguen, Alex Sergeev, Ben Kadlec, Rosanne Liu, and Jason Yosinski.
\newblock Faster neural networks straight from jpeg.
\newblock In S. Bengio, H. Wallach, H. Larochelle, K. Grauman, N. Cesa-Bianchi,
  and R. Garnett, editors, {\em Advances in Neural Information Processing
  Systems}, volume~31. Curran Associates, Inc., 2018.

\bibitem{he2016deep}
Kaiming He, Xiangyu Zhang, Shaoqing Ren, and Jian Sun.
\newblock Deep residual learning for image recognition.
\newblock In {\em Proceedings of the IEEE conference on computer vision and
  pattern recognition}, pages 770--778, 2016.

\bibitem{horn2017design}
Daniel~Reiter Horn, Ken Elkabany, Chris Lesniewski-Lass, and Keith Winstein.
\newblock The design, implementation, and deployment of a system to
  transparently compress hundreds of petabytes of image files for a
  file-storage service.
\newblock In {\em 14th USENIX Symposium on Networked Systems Design and
  Implementation (NSDI 17)}, pages 1--15, 2017.

\bibitem{kingma2014adam}
Diederik~P Kingma and Jimmy Ba.
\newblock Adam: A method for stochastic optimization.
\newblock {\em arXiv preprint arXiv:1412.6980}, 2014.

\bibitem{kingma2019bit}
Friso~H Kingma, Pieter Abbeel, and Jonathan Ho.
\newblock Bit-swap: Recursive bits-back coding for lossless compression with
  hierarchical latent variables.
\newblock {\em arXiv preprint arXiv:1905.06845}, 2019.

\bibitem{li2018learning}
M. Li, W. Zuo, S. Gu, D. Zhao, and D. Zhang.
\newblock Learning convolutional networks for content-weighted image
  compression.
\newblock In {\em 2018 IEEE/CVF Conference on Computer Vision and Pattern
  Recognition}, pages 3214--3223, Salt Lake City, UT, USA, June 2018.

\bibitem{li2020optimizing}
Zhijing Li, Christopher De~Sa, and Adrian Sampson.
\newblock Optimizing jpeg quantization for classification networks.
\newblock In {\em Resource-Constrained Machine Learning (ReCoML) Workshop of
  MLSys 2020 Conference}, Austin, TX, USA, 2020.

\bibitem{liu2018deepn}
Z. {Liu}, T. {Liu}, W. {Wen}, L. {Jiang}, J. {Xu}, Y. {Wang}, and G. {Quan}.
\newblock Deepn-jpeg: A deep neural network favorable jpeg-based image
  compression framework, 2018.

\bibitem{mentzer2019practical}
Fabian Mentzer, Eirikur Agustsson, Michael Tschannen, Radu Timofte, and Luc~Van
  Gool.
\newblock Practical full resolution learned lossless image compression.
\newblock In {\em Proceedings of the IEEE Conference on Computer Vision and
  Pattern Recognition}, pages 10629--10638, 2019.

\bibitem{mentzer2018conditional}
F. Mentzer, E. Agustsson, M. Tschannen, R. Timofte, and L. Van~Gool.
\newblock Conditional probability models for deep image compression.
\newblock In {\em 2018 IEEE/CVF Conference on Computer Vision and Pattern
  Recognition}, pages 4394--4402, Salt Lake City, UT, USA, June 2018.

\bibitem{mentzer2020learning}
Fabian Mentzer, Luc~Van Gool, and Michael Tschannen.
\newblock Learning better lossless compression using lossy compression.
\newblock In {\em Proceedings of the IEEE/CVF Conference on Computer Vision and
  Pattern Recognition}, pages 6638--6647, 2020.

\bibitem{minnen2018joint}
D. Minnen, J. Ball{\'e}, and G.~D. Toderici.
\newblock Joint autoregressive and hierarchical priors for learned image
  compression.
\newblock In {\em Advances in Neural Information Processing Systems 31}, pages
  10771--10780, Montreal, QC, Canada, December 2018.

\bibitem{Authors14}
Full~Author Name.
\newblock The frobnicatable foo filter, 2014.
\newblock Face and Gesture submission ID 324. Supplied as additional material
  {\tt fg324.pdf}.

\bibitem{Authors14b}
Full~Author Name.
\newblock Frobnication tutorial, 2014.
\newblock Supplied as additional material {\tt tr.pdf}.

\bibitem{pavlov20197z}
Igor Pavlov.
\newblock 7z format.
\newblock {\em \url{http://www.7-zip.org/7z.html}}, 2019.

\bibitem{qiu2020deep}
Han Qiu, Qinkai Zheng, Gerard Memmi, Jialiang Lu, Meikang Qiu, and Bhavani
  Thuraisingham.
\newblock Deep residual learning-based enhanced jpeg compression in the
  internet of things.
\newblock {\em IEEE Transactions on Industrial Informatics}, 17(3):2124--2133,
  2020.

\bibitem{rabbani2002jpeg2000}
Majid Rabbani.
\newblock Jpeg2000: Image compression fundamentals, standards and practice.
\newblock {\em Journal of Electronic Imaging}, 11(2):286, 2002.

\bibitem{rippel2017real}
O. Rippel and L. Bourdev.
\newblock Real-time adaptive image compression.
\newblock In {\em Proceedings of the 34th International Conference on Machine
  Learning}, pages 2922--2930, Sydney, NSW, Australia, August 2017.

\bibitem{sikora1997mpeg}
Thomas Sikora.
\newblock Mpeg digital video-coding standards.
\newblock {\em IEEE signal processing magazine}, 14(5):82--100, 1997.

\bibitem{theis2017lossy}
L. Theis, W. Shi, A. Cunningham, and F. Husz{\'a}r.
\newblock Lossy image compression with compressive autoencoders.
\newblock In {\em Proceedings of the 5th International Conference on Learning
  Representations}, Toulon, France, April 2017.

\bibitem{toderici2015variable}
G. Toderici et~al.
\newblock Variable rate image compression with recurrent neural networks.
\newblock In {\em Proceedings of the 4th International Conference on Learning
  Representations}, San Juan, Puerto Rico, May 2016.

\bibitem{van2016conditional}
A. van~den Oord et~al.
\newblock Conditional image generation with {PixelCNN} decoders.
\newblock In {\em Advances in Neural Information Processing Systems 29}, pages
  4790--4798, Barcelona, Spain, December 2016.

\bibitem{oord2016pixel}
A. van~den Oord, N. Kalchbrenner, and K. Kavukcuoglu.
\newblock Pixel recurrent neural networks.
\newblock In {\em Proceedings of the 33rd International Conference on Machine
  Learning}, pages 1747--1756, New York, NY, USA, June 2016.

\bibitem{wallace1991the}
Gregory~K Wallace.
\newblock The jpeg still picture compression standard.
\newblock {\em Communications of The ACM}, 34(4):30--44, 1991.

\bibitem{learningfrequency}
Kai Xu, Minghai Qin, Fei Sun, Yuhao Wang, Yen-kuang Chen, and Fengbo Ren.
\newblock Learning in the frequency domain.
\newblock {\em arXiv preprint arXiv:2002.12416}, 2020.

\bibitem{yuan2020image}
Xin Yuan and Raziel Haimi-Cohen.
\newblock Image compression based on compressive sensing: End-to-end comparison
  with jpeg.
\newblock {\em IEEE Transactions on Multimedia}, 22(11):2889--2904, 2020.

\end{thebibliography}
}

\end{document}